\documentclass[runningheads]{llncs}
\usepackage[T1]{fontenc}
\usepackage{graphicx}
\usepackage{wrapfig}
\usepackage{graphicx} 

\usepackage{amsmath}
\usepackage{amssymb}

\begin{document}
\title{Score Matching on Large Geometric Graphs for Cosmology Generation}
%
%
\author{Diana-Alexandra Onuțu\inst{1,2\thanks{Work is partially done during internship in High-Performance ML Team at SURF. \mailname\ d.onutu@tue.nl}}\orcidID{0009-0009-3933-5071} \and
Yue Zhao\inst{2}\orcidID{0009-0005-4074-5116} \and
Joaquin Vanschoren\inst{1}\orcidID{0000-0001-7044-9805} \and Vlado Menkovski\inst{1}\orcidID{0000-0001-5262-0605}}
\authorrunning{D. Onuțu et al.}
%
\institute{Eindhoven University of Technology, Eindhoven, The Netherlands\\
\email{\{d.onutu, j.vanschoren, v.menkovski\}@tue.nl}\\ \and
SURF, Amsterdam, The Netherlands\\
\email{yue.zhao@surf.nl}}
\maketitle              
\begin{abstract}

Generative models are a promising tool to produce cosmological simulations but face significant challenges in scalability, physical consistency, and adherence to domain symmetries, limiting their utility as alternatives to $N$-body simulations. To address these limitations, we introduce a score-based generative model with an equivariant graph neural network that simulates gravitational clustering of galaxies across cosmologies starting from an informed prior, respects periodic boundaries, and scales to full galaxy counts in simulations. A novel topology-aware noise schedule, crucial for large geometric graphs, is introduced. The proposed equivariant score-based model successfully generates full-scale cosmological point clouds of up to 600,000 halos, respects periodicity and a uniform prior, and outperforms existing diffusion models in capturing clustering statistics while offering significant computational advantages. This work advances cosmology by introducing a generative model designed to closely resemble the underlying gravitational clustering of structure formation, moving closer to physically realistic and efficient simulators for the evolution of large-scale structures in the universe.

\end{abstract}

\section{Introduction}

Influenced by gravity, a nearly uniform distribution of matter evolves into a complex network of clusters, filaments, and voids, known as the large-scale structure of the universe. Analyzing this structure provides valuable information for inferring properties of dark matter and dark energy, advancing our understanding of cosmic structure formation and fundamental physics. Cosmologists study this by comparing observational data collected using telescopes with predictions from theoretical models, such as those produced with $N$-body simulations. These comparisons rely on summary statistics that quantify and describe the matter distribution, with the aim of testing and constraining cosmological theories.

One of the leading models for the large-scale structure of the universe is the Lambda Cold Dark Matter ($\Lambda$CDM) model. Based on this model, $N$-body simulations track the dynamics of dark matter particles over cosmic time, each influenced by mutual gravitational interactions and governed by $\Lambda$CDM parameters. The simulations result in particle distributions that form the complex large-scale structure characterized by clusters of matter and vast voids. Within these distributions, dark matter particles collapse under gravity to form gravitationally bound structures known as halos, sites for galaxy formation. As such, the spatial distribution of dark matter halos serves as a theoretical counterpart to the observed galaxy distributions. These observations are then compared with galaxy surveys, allowing cosmologists to test the consistency of $\Lambda$CDM and extract constraints on model parameters governing the formation of cosmic structures.

Ideally, a large number of high-resolution $N$-body simulations are required to comprehensively explore the cosmological parameter space, but their high computational cost makes this challenging. For example, Millennium Simulation II~\cite{boylan2009resolving}, which evolved $\sim$ 10 billion dark matter particles, required approximately one month and 1.4 million CPU hours, while a higher-resolution simulation of $\sim$70 billion particles~\cite{teyssier2009full} ran for two months, and the Quijote simulation suite~\cite{villaescusa2020quijote}, consisting of 44,100 simulations with millions of particles each, required more than 35 million CPU hours.

To overcome the computational challenges of $N$-body simulations, generative models provide a promising alternative to significantly improve the computational complexity~\cite{cuesta2023point,riveros2025conditional}. These methods adopt a data-driven approach to approximate the simulation process. Specifically, given a set of cosmologies, the machine learning-based (ML) models estimate the arrangement of dark matter in space conditioned on cosmological parameters.

A recent approach introduces a diffusion-based generative model to generate samples of large-scale structure based on halo distributions~\cite{cuesta2023point}. To mitigate information loss associated with discretized representations such as pixels or voxels, the method uses point clouds limited to the 5,000 most massive halos per cosmology. The model architecture is based on graph neural networks (GNN) or Transformers to account for permutation invariance of point clouds. This approach shows promising results in generating samples that accurately reproduce summary statistics across small to intermediate scales.

However, the previous approach presents several limitations. First, it initializes sampling from a Gaussian prior, which does not reflect the near-uniform matter distribution characteristic of the early universe. This mismatch can make the modeling task more difficult and inefficient. Additionally, the Gaussian prior is defined on an unbounded domain, while cosmological simulations are confined to a bounded, periodic volume. The model must therefore implicitly map unbounded inputs to the bounded simulation space in the absence of an explicit mechanism enforcing spatial boundary constraints. Second, the model does not explicitly encode key domain symmetries: cosmological data are equivariant under the $E(3)$ Euclidean group and exhibit periodic boundary conditions (PBC). Incorporating such symmetries as inductive biases is known to enhance model generalization, as demonstrated by the success of translation-equivariant architectures such as convolutional neural networks. Finally, generated point clouds are limited to only about 1\% of the number of halos in simulations, significantly fewer than the number of halos targeted in full simulations. Overall, these limitations hinder the scalability and physical fidelity required for generative models to serve as viable surrogates for computationally expensive $N$-body simulations.

This work addresses the aforementioned challenges by introducing a score-based generative model~\cite{song2019generative} to sample large-scale structure conditioned on cosmological parameters. Following~\cite{cuesta2023point}, we represent the matter distribution as point clouds and construct graphs based on $k$-nearest neighbors ($k$-NN). A key difference from diffusion models lies in the noise schedule: instead of interpolating between the data distribution and a Gaussian, random noise is added to the data. This results in the most corrupted state corresponding to a uniform halo distribution, enabling more efficient sampling from an informed prior. PBCs are imposed as in~\cite{xiecrystal} to align the bounded, periodic domain of cosmological simulations. To respect Euclidean symmetries, we employ an $E(3)$ equivariant graph neural network (EGNN)~\cite{satorras2021n}. By incorporating these domain-specific constraints: periodic boundaries, symmetries, and uniform prior, the model's denoising can be interpreted as a simulator of gravitational clustering, providing a physically grounded generative framework for structure formation.

Developing score-based generative models for cosmology involves two main challenges. First, the noise schedule must be carefully designed to map an initially uniform point distribution into the highly clustered halos of large-scale structure. Second, the model must scale to generate point clouds with up to 600,000 halos to capture the full complexity of gravitational clustering. To address the first challenge, we introduce a method for quantifying changes in graph topology across noise levels, providing a principled approach to tuning the noise schedule. To address scalability, we train on smaller sub-volumes sampled from the full simulation box and perform full-resolution generation at inference time.

Prior work~\cite{song2020improved} shows score-based generative dependence on noise schedule design, with performance degrading when scaling from low- to high-resolution images. In the context of graphs, diffusion and score-based models have been mostly limited to small graphs with 4 to 400 nodes~\cite{jo2022score,niu2020permutation,xiecrystal}, and generation quality degrades noticeably for larger graphs~\cite{jo2022score}. For point clouds, existing models typically operate at scales of 1,000 and 2,048 points~\cite{luo2021score,luo2021diffusion}. Within cosmology, where datasets naturally consist of large point clouds, generative approaches have been scaled from 100 points~\cite{makinen2022cosmic} up to 5,000 points~\cite{cuesta2023point}, highlighting current limitations in generating full-scale point clouds for structure generation.

In this work, we introduce a score-based generative model to sample cosmologies. We summarize our contributions below:
\begin{itemize}
    \item We introduce a score-based model for cosmologies that generates matter distribution across various $\Lambda$CDM configurations, enforcing periodic boundary conditions, and initializing sampling from a physically motivated uniform prior, thereby aligning the generative process more closely with physical reality compared to Gaussian-initialized diffusion models.
    \item We incorporate domain symmetries via an equivariant graph neural network, ensuring consistency with cosmic structure and enhancing model generalization and data efficiency.
    \item We introduce a principled, topology-aware method for designing noise schedules critical for large geometric graphs, addressing a key challenge in scaling score-based models. This is vital because the structure of large graphs is highly sensitive to noise.
    \item The unique combination of $E(3)$ equivariance, PBCs, physically motivated prior, and a scalable noise schedule tailored for large geometric graphs represents a significant step forward in cosmology simulation, outperforming prior diffusion models. This synergy allows the model to not only generate samples, but do so in a way that reflects the underlying physical processes of structure formation more faithfully than previous approaches.
   
\end{itemize}

\section{Cosmology representation}
$N$-body simulations model the gravitational evolution of a large number of dark matter particles from early-universe initial conditions to present epoch (redshift $z=0$). These simulations run within a periodic box, where particles exiting one side re-enter from the opposite side, mimicking the infinite nature of the universe. Particles interact through gravity, leading to the emergence of non-linear structures from an initially near-uniform density field. The resulting matter distribution reflects the large-scale structure of the universe under a specific $\Lambda$CDM model configuration. Thus, each simulation represents a possible instance of the universe consistent with the assumed model parameters, namely, a cosmology.

\begin{table}[h!]
\centering
\caption{Parameters of the $\Lambda$CDM model and corresponding value ranges used in the Latin hypercube sampling of the Quijote simulations \cite{villaescusa2020quijote}.}
\label{tab:cosmo_params}
\begin{tabular}{c c c} 
\hline
 Parameter & Description & Range \\
 \hline
 $\Omega_m$  & Total matter density & $[0.1, 0.5]$    \\ 
 $\Omega_b$  & Baryon matter density & $[0.03, 0.07]$  \\ 
 $H_0 $        & Hubble constant & $[0.5, 0.9]$    \\
 $n_s$       & Scalar spectrum power-law index & $[0.8, 1.2]$   \\
 $\sigma_8$  & Amplitude of density fluctuations at scale $8$ $h^{-1}$ Mpc &$[0.6, 1.0]$    \\
 \hline
\end{tabular}
\end{table}

\begin{figure}[h!]
    \centering
    \includegraphics[scale=0.5]{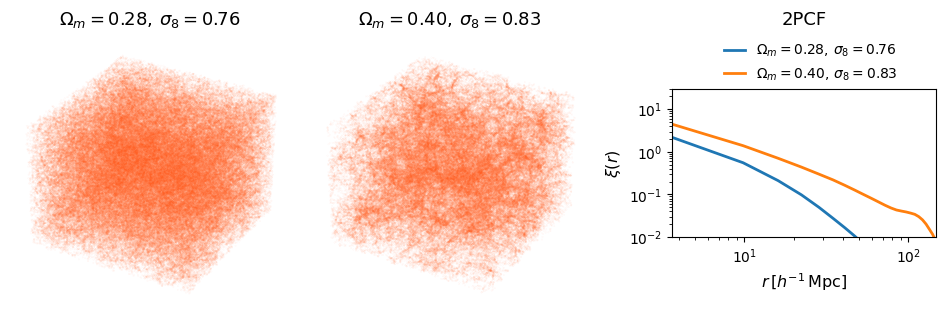}
    \caption{Large-scale structures and 2PCF of two cosmologies varying ($\Omega_m$, $\sigma_8$), drawn from the standard Latin-hypercube sampling at $z=0$ in the Quijote simulations~\cite{villaescusa2020quijote}.}
    \label{fig:particles_cosm_params}
\end{figure}

Each cosmology is characterized by a distinct set of $\Lambda$CDM parameter values, described in Table \ref{tab:cosmo_params}, which influence the formation of large-scale structure. Among these, $\Omega_m$ quantifies the total matter content while $\sigma_8$ characterizes the clumpiness factor at the 8 $h^{-1}$ megaparsec (Mpc) scale. The information content of cosmologies is quantified through summary statistics. The two-point correlation function (2PCF), denoted $\xi(r)$, measures clustering strength by quantifying the excess probability, relative to a uniform distribution, of finding pairs of points separated by distance $r$. Higher 2PCF values indicate stronger clustering; lower values suggest uniformity. Figure \ref{fig:particles_cosm_params} shows the resulting large-scale structure and 2PCF for various $(\Omega_m, \sigma_8)$ combinations. Notably, high $\Omega_m$ and high $\sigma_8$ yield a highly clustered distribution with prominent voids, whereas low $\Omega_m$ and low $\sigma_8$ produce a more uniform and weakly clustered matter distribution.

\section{Methodology}
We propose a score-based generative model as an alternative to the diffusion-based approach used in~\cite{cuesta2023point} for generating cosmological samples. This formulation offers two key advantages: (i) it enables the explicit enforcement of periodic boundary conditions during both training and inference, and (ii) it adopts a uniform prior, more consistent with the nearly homogeneous matter distribution in the early universe than the commonly used Gaussian prior. Formally, given a set of halos $\mathbf{X}$, representing their spatial positions, the objective is to model the conditional distribution $p(\mathbf{X}|\vec{\theta})$, where $\vec{\theta}$ denotes the cosmological parameters.

Score-based~\cite{song2019generative} and diffusion~\cite{ho2020denoising} models follow a similar framework: a forward process $q(\tilde{\mathbf{X}}|\mathbf{X})$ that corrupts data into a noisy version $\tilde{\mathbf{X}}$, and a reverse process where a model $p_\theta(\mathbf{X})$ is trained to iteratively denoise $\tilde{\mathbf{X}}$, generating samples from the data distribution $p(\mathbf{X})$. The forward process follows a noise schedule that progressively perturbs the signal until it becomes pure noise. For matter distribution, this perturbation involves progressively displacing halos from their initial position, leading to a changed large-scale structure. Conversely, the denoising process aims to restore halos to their original position, thereby generating the large-scale structure of the cosmology. To realistically simulate structure formation, the generative model must ensure that halos remain within the simulation bounds, re-entering from the opposite side when displaced beyond a limit.

\begin{figure}[h!]
    \centering
    \includegraphics{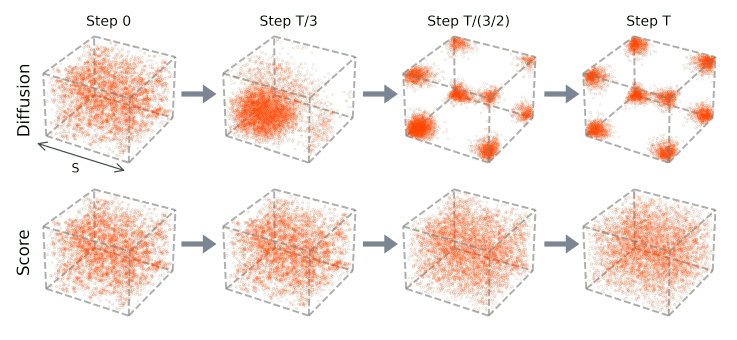}
    \caption{Forward process visualization of diffusion and score-based models with periodic boundaries. Diffusion models exhibit artificial clustering near boundaries due to wrapping, while score-based models produce a uniform distribution across the domain.}
    \label{fig:perturbation_pbc}
\end{figure}

A subtle difference in the noise schedules makes diffusion models less suitable for periodically bounded data. Figure \ref{fig:perturbation_pbc} shows the forward process over $T$ steps for both models, incorporating periodic wrapping at box length $S$. In diffusion models, the noise schedule gradually transforms the data into a Gaussian, requiring the denoising process to recover the large-scale structure from this prior. Thus, the model does not simulate the clustering of dark matter halos from a physically realistic initial condition. The forward process with wrapping is defined as an interpolation between the data and noise distribution: $\tilde{\textbf{X}}_t = (\sqrt{\bar{\alpha}_t} \cdot \textbf{X} + \sqrt{1- \bar{\alpha}_t} \cdot \vec{\epsilon}) \bmod S$, where the coefficients balance data and noise at timestep $t$, and $\vec{\epsilon} \sim \mathcal{N}(0, I)$. The modulo operation keeps halos within the box by wrapping values outside the boundaries back into the domain. As $t$ increases, the signal diminishes, and data approaches a Gaussian centered at 0 when the variance is small. This wrapping effect causes artificial clustering at the boundaries, producing an unrealistic cosmological state.

In contrast, score-based models perturb data by displacing halos with some noise. The forward process under periodic boundaries is defined as: $\tilde{\mathbf{X}}_\sigma = (\mathbf{X} + \sigma \cdot \vec{\epsilon}) \bmod S$, where $\vec{\epsilon} \sim \mathcal{N}(0, I)$ and $\sigma$ controls the noise level. As $\sigma$ increases, halos move further away, and at sufficiently high noise, the perturbed data approximate a uniform distribution due to periodic wrapping, avoiding the boundary clustering seen in diffusion models. Thus, the score-based model must learn an implicit representation of matter clustering to generate large-scale structure.

Next, we formally define the modeling task using score estimation, following the formulation in \cite{song2019generative,song2020improved}, and enforcing periodic boundary conditions as described in \cite{xiecrystal}. The generative process is defined through the estimation of the \textit{score function}, which is the gradient of the log-probability density of the data corrupted with Gaussian noise. Therefore, the score estimates the direction in which the perturbed data should be moved to recover the data distribution. 

Given a set of $N$ halos, each with 3D coordinates, their positions are represented as $\mathbf{X} \in [0, S)^{N \times 3}$, where $S$ is the box length. Let $q_\sigma(\tilde{\mathbf{X}} | \mathbf{X})$ be a noise distribution  with periodic wrapping, where $\{\sigma_i\}_{i=1}^L$ is a geometric sequence of $L$ positive noise levels. The objective is to learn the score function $\nabla_{\mathbf{X}} \log q_\sigma(\mathbf{X})$. Since $ q_\sigma(\tilde{\mathbf{X}} | \mathbf{X}) = \mathcal{N}(\tilde{\mathbf{X}} | \mathbf{X}, \sigma^2, I)$, we have $\nabla_{\tilde{\mathbf{X}}} \log q_\sigma(\tilde{\mathbf{X}} | \mathbf{X}) = -\vec{d}_{\text{min}}(\tilde{\mathbf{X}}, \mathbf{X}) / \sigma^2$, where $\vec{d}_{\text{min}} = (\tilde{\mathbf{X}} - \mathbf{X}) \bmod S$, following~\cite{xiecrystal}.

A conditional score network $\mathbf{s}_{\vec{\theta}}$ is trained to estimate gradients of the log-density for all noise levels, satisfying: $\mathbf{s}_{\vec{\theta}}(\tilde{\mathbf{X}}, \sigma) \approx \nabla_{\tilde{\mathbf{X}}} \log q_\sigma(\tilde{\mathbf{X}} | \mathbf{X})$. The training objective is \textbf{denoising score matching}, defined for a specific noise level $\sigma$ as:

\begin{equation}
    \ell(\vec{\theta}; \sigma) \triangleq \frac{1}{2} \mathbb{E}_{p_{\text{data}}(\mathbf{X})} \mathbb{E}_{\tilde{\mathbf{X}} \sim \mathcal{N}(\mathbf{X}, \sigma^2 I)} \left[ \left\| 
    \frac{\mathbf{s}_{\boldsymbol{\theta}} (\tilde{\mathbf{X}})}{\sigma} + \frac{\vec{d}_{\text{min}}(\tilde{\mathbf{X}}, \mathbf{X})}{\sigma^2} \right\|_2^2 \right],
\end{equation}

conditioning $\mathbf{s}_{\vec{\theta}}$ on $\sigma$ by scaling its prediction, avoiding the need for explicit noise level conditioning parameters as used in diffusion models. The overall loss aggregates the per-noise objectives: $\mathcal{L} \left( \boldsymbol{\theta}; \{\sigma_i\}_{i=1}^{L} \right) \triangleq \frac{1}{L} \sum_{i=1}^{L} \sigma_i^2 \ell (\boldsymbol{\theta}; \sigma_i)$. Furthermore, sampling is performed using annealed Langevin dynamics, as described in~\cite{song2019generative}, with PBCs and $\alpha_i$ learning rate applied at each update step:

\begin{equation}
    \tilde{\mathbf{X}}_t \gets (\tilde{\mathbf{X}}_{t-1} + \frac{\alpha_i}{2} s_\theta (\tilde{\mathbf{X}}_{t-1}, \sigma_i) + \sqrt{\alpha_i} \mathbf{z}_t ) \bmod S, \quad \quad\mathbf{z}_t \sim \mathcal{N}(0, I).
\end{equation}

\subsection{Score Matching on Large Geometric Graphs}

Noise levels $\{\sigma_i\}_{i=1}^L$ must be chosen such that $\sigma_L$ is small enough to closely approximate the true data distribution, while $\sigma_1$ is large enough to enable score estimation in low-density regions, as discussed in~\cite{song2019generative}. For image data, these levels are often selected via visual inspection: low noise level preserves most of the image, while high noise produces unrecognizable images.

However, for graph-structured data, such heuristic approaches prove inadequate. As such, our main aim is to directly address the challenge of systematically defining a noise schedule suitable for graphs of arbitrary size. This requires a principled notion of structural change and a metric to quantify the deviation of a perturbed graph from its original form. Motivated by the insight from~\cite{song2020improved} that noise schedule affects generative performance in high-resolution images, we propose a method to quantify change in graph topology to define noise levels. We define the similarity between a perturbed and original graph in terms of topological structure, measured by the percentage of preserved edges. Specifically, we compute the proportion of shared $k$-nearest neighbor connections between the original and perturbed graphs across various noise levels. This approach allows for consistent and scalable noise scheduling across diverse graph sizes.

\subsection{Equivariant Graph Neural Networks for Cosmology Modeling}
The score network employs a GNN to process the perturbed graphs and estimate scores. We adopt the message-passing framework \cite{battaglia2018relational,gilmer2017neural}, leveraging the $E(n)$ equivariant graph neural network~\cite{satorras2021n}, with modifications tailored to our task. 

Given node features $\mathbf{h}_i \in \bbbr^n$, coordinates $\mathbf{x}_i \in [0,S)^3$, edges $e_{ij} \in \mathcal{E}$, scores $\mathbf{s}_i \in \bbbr^3$ (initialized at zero), and global graph conditioning $\vec{\theta} \in \bbbr^d$, where $n$ is the feature dimension and $d$ is the number of cosmological parameters, the equivariant graph convolutional layer (EGCL) is defined as: $\mathbf{h}^{l+1}, \mathbf{s}^{l+1} = \text{EGCL}[\mathbf{h}^l, \mathbf{s}^l, \mathbf{x}, \mathcal{E}, \vec{\theta}]$, where $l$ refers to the messaging passing layer.  We retain the original message, aggregation and feature update functions from~\cite{satorras2021n}, but change their input: add global conditioning $\vec{\theta}$ and use relative differences in features and coordinates, as it improves training compared to concatenation or Euclidean distances. Unlike the original model, which predicts coordinates, we predict scores as follows:

\begin{equation}
    \mathbf{s}_i^{l+1} = \mathbf{s}_i^l + \frac{1}{|\mathcal{N}_i|} \sum_{j \in \mathcal{N}_i} (\mathbf{x}_i - \mathbf{x}_j) \phi_x(\mathbf{m}_{ij}),\quad \mathcal{N}_i \text{ is neighborhood of node } v_i.
\end{equation}

The scores are iteratively refined through message-passing, progressively incorporating information from a larger neighborhood. This separation between coordinates and predicted scores ensures that scores remain defined in physical space while preserving all necessary symmetries. To enforce global translation invariance, we adopt the method of \cite{garcia2021n}, subtracting the center of gravity from the predicted scores. This constrains the predictions to a translation-invariant subspace by ensuring a zero mean. As a result, the model's output depends only on the relative arrangement of the points, independent of the absolute position of the system in space. Implementation details are provided in Appendix~\ref{appendix}.

\section{Experiments}

We conduct experiments using halo catalogs from the standard regular-resolution Latin Hypercube at redshift $z=0$, part of the Quijote $N$-body simulations suite~\cite{villaescusa2020quijote}. The dataset includes 2,000 simulated halo catalogs varying configurations of the $\Lambda$CDM model, as shown in Table~\ref{tab:cosmo_params}. Each simulation is initialized with a unique random seed and runs within a fixed periodic volume $(h^{-1}\text{Gpc})^3$.

For evaluation, we use the 5,000 most massive halos per simulation from~\cite{cuesta2023point} to ensure a fair comparison. The dataset includes 1,800 cosmologies for training and 200 for testing. For large graphs training and generation, we use the full halo catalogs from~\cite{villaescusa2020quijote}. Each halo catalog represents a cosmology and is treated as a point cloud. The 3D coordinates are normalized to $[0, 1]$ for training. To model the spatial structure, each point cloud is converted into a graph using $k$-NN with $k = 20$, incorporating PBCs via the Minimum Image Convention. This construction preserves local geometry while respecting the periodicity of the simulation volume.

\subsection{Noise Schedule}
\begin{figure}[h!]
    \centering
    \includegraphics{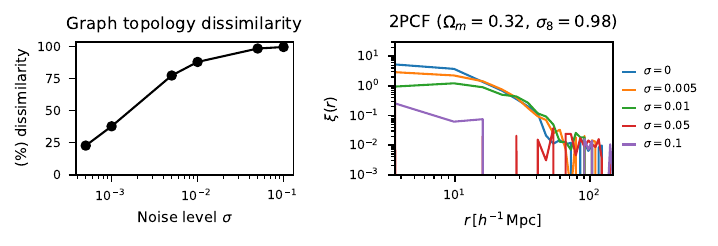}
    \caption{Effect of noise on graph topology and clustering strength (via 2PCF).}
    \label{fig:noise_analysis_combined}
\end{figure}

To design the noise schedule, we quantify topological change by the percentage of preserved edges under perturbation. Specifically, we compute the average edge difference across all cosmologies for varying noise levels $\sigma$ and the impact of $\sigma$ on 2PCF a random cosmology. Results are illustrated in Figure \ref{fig:noise_analysis_combined}. The commonly used setting $\sigma_L = 0.01$ causes over 80\% edge difference on average and low 2PCF values, indicating that even low noise substantially alters graph topology and clustering strength. This is problematic for score-based models, which require the perturbed graphs at low noise to remain close to the data distribution. The sensitivity arises from graph size: in large graphs, even minor perturbations can significantly disrupt local neighborhoods, as the $k$-NN structure is highly sensitive to small displacements. In contrast, smaller graphs are more robust to the same level of noise, which may explain why previous work has not required a systematic approach to noise scheduling. Motivated by this analysis, we adopt $\sigma_L = 0.001$, which preserves roughly 30\% of the original edges and yields a closer match to the clustering strength of the underlying cosmology. Additionally, we set the maximum noise level to $\sigma_1=0.2$, where the graph's topology and clustering strength are degraded, ensuring complete signal removal.

\subsection{Quantitative Analysis}
We evaluate diffusion and (non)equivariant score-based generative models for cosmology generation. One model is conditioned on all cosmological parameters $\vec{\theta} = (\Omega_m, \Omega_b, H_0, \sigma_8, n_s)$, the other ones only on $(\Omega_m, \sigma_8)$. We also investigate the impact of a different noise schedule. Performance is evaluated by calculating the mean absolute error (MAE) and mean squared error (MSE) of the 2PCF over 10 generated samples for each of the 200 validation cosmologies. Results are then averaged over all scales and reported in Table \ref{tab:model_performance}.

\begin{table}[h!]
    \centering
    \caption{Overall model performance comparison. ScoreEGNN ($\Omega_m$, $\sigma_8$)** uses a noise schedule commonly found in literature. We trained the diffusion models from~\cite{cuesta2023point}.}
    \label{tab:model_performance}
    \begin{tabular}{ lcc }
    \hline
        Model  & MAE 2PCF $\downarrow$ & MSE 2PCF $\downarrow$ \\  
        \hline
        DiffusionGNN ($\Omega_m$, $\sigma_8$) & 0.4625 & 2.7809 \\  
        DiffusionTransformer ($\Omega_m$, $\sigma_8$)  & 0.2297 & 0.7147 \\
        ScoreGNN ($\Omega_m$, $\sigma_8$) & 0.2366 &  0.7092 \\
        ScoreEGNN ($\Omega_m$, $\sigma_8$) & 0.2250 & 0.6638 \\
        ScoreEGNN $\vec{\theta}$ & \textbf{0.1764} & \textbf{0.3837} \\
        ScoreEGNN ($\Omega_m$, $\sigma_8$)** & 0.8606 & 7.2882  \\
        \hline
    \end{tabular}
\end{table}

Conditioning on the full parameter set yields the best performance, highlighting the importance of incorporating domain-specific information. Equivariance provides only marginal improvements over the data-augmented non-equivariant ScoreGNN. This can be explained by two factors. First, the 2PCF metric only allows for limited discrimination among cosmology structures, masking some of the advantages of an equivariant model. Second, the ScoreGNN model is able to learn symmetry properties implicitly. Naive noise schedules significantly degrade generative quality, as expected. Score-based (equivariant) GNNs outperform DiffusionGNN, while DiffusionTransformer shows comparable performance, indicating that the Transformer architecture may have sufficient capacity to model the large-scale structure without explicit inductive biases.

\subsection{Qualitative Analysis}
\label{quali}
\begin{figure}[h!]
    \centering
    \includegraphics{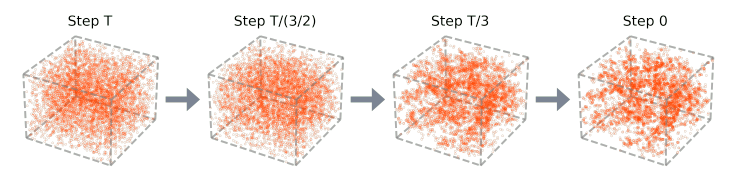}
    \caption{ScoreEGNN($\Omega_m$, $\sigma_8$) sampling process for $(\Omega_m = 0.10$, $\sigma_8 = 0.92)$. Generated sample presents dense regions and voids, as expected for this configuration.}
    \label{fig:sampling}
\end{figure}

\begin{wrapfigure}{r}{0.5\textwidth}
    \vspace{-2em}
    \centering
    \includegraphics[width=0.48\textwidth]{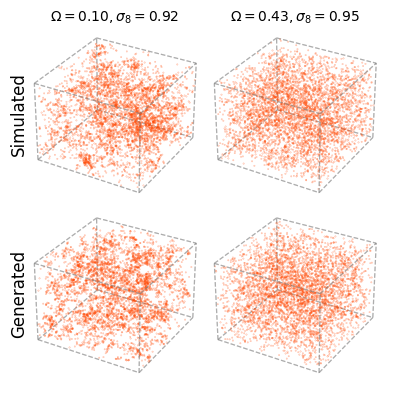}
    \caption{Simulated and generated cosmologies. Cosmological dependence is evident. High-density regions indicate strong clustering; sparse regions correspond to voids.}
    \label{fig:generated_samples}
    \vspace{-3em}
\end{wrapfigure}

As shown in Figure \ref{fig:sampling}, the sampling process begins from a uniform prior and progressively evolves into a clustered large-scale structure, indicating that the model estimates scores that effectively guide halo dynamics. Additionally, Figure \ref{fig:generated_samples} demonstrates that the model captures the dependence on cosmological parameters, reproducing the trend observed in the simulations.

\begin{figure}
    \centering
    \includegraphics{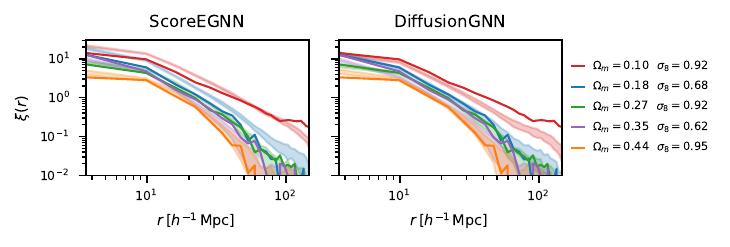}
    \caption{2PCF for 5 $\Omega_m$ equally spaced cosmologies. Solid line represents the 2PCF of the simulated cosmologies, while the contour is based on the mean and standard deviation of 20 generated samples per configuration.}
    \label{fig:quali_tpcf}
\end{figure}

Furthermore, in Figure \ref{fig:quali_tpcf} we evaluate the clustering strength of generated samples with both the ScoreEGNN and DiffusionGNN models, conditioned on ($\Omega_m$, $\sigma_8$). Both models generally reproduce the overall trend of the simulated data 2PCF. However, both models generate samples with high clustering variance at large scales, consistent with the findings of~\cite{cuesta2023point}, indicating that both generative models struggle to reproduce the clustering patterns on large scales. This reinforces the idea that the problem may be the graph representation and message-passing mechanism, known to struggle with long-range correlations.

\subsection{Efficiency Analysis}

We also assess computational efficiency. Table~\ref{tab: model_efficieny} shows score-based models outperform diffusion models across all metrics. Score-based models generate the same number of samples roughly twice as fast as diffusion models. Reducing the computational cost of $N$-body simulations is a key motivation in cosmology. Our model generates 2,000 halo catalogs in approximately one hour on 1 H100 GPU. In contrast, the Quijote suite simulations require an average of 1.6 million CPU hours for the same number of halo catalogs. This is a speedup of more than six orders of magnitude.

\begin{table}[h!]
\centering
\caption{Model efficiency comparison. Inference time is time to generate 2,000 samples.}
\label{tab: model_efficieny}
\begin{tabular}{ lcccc}
\hline
 Model  & \# parameters & \# iterations & Training time & Inference time \\ 
 \hline
 DiffusionGNN & 629K & 192K & 12h & 3h30min \\  
 DiffusionTransformer & 4.8M & 265K & 12h  & 2h30min\\
 ScoreGNN & 570K & 28K  & 5h30min & 1h5min\\
 ScoreEGNN & 417K & 28K & 5h45min & 1h10min \\
 \hline
\end{tabular}
\end{table}

\subsection{Large Graphs Generation}

Previously, we generated cosmologies containing only the 5,000 most massive halos, omitting the vast majority of halos in a catalog. In this section, we scale the ScoreEGNN ($\Omega_m$, $\sigma_8$) model to generate cosmologies with up to 600,000 halos. Training on full graphs of this scale is computationally prohibitive, primarily due to the time and memory cost of our custom PBC-preserving $k$-NN computations. To mitigate this, we train on smaller sub-volumes obtained via box subsampling. During inference, we operate on the full graph. This means we can only generate one cosmology at a time on a single GPU due to memory limitations. 

\begin{figure}[h!]
    \centering
    \includegraphics[scale=0.6]{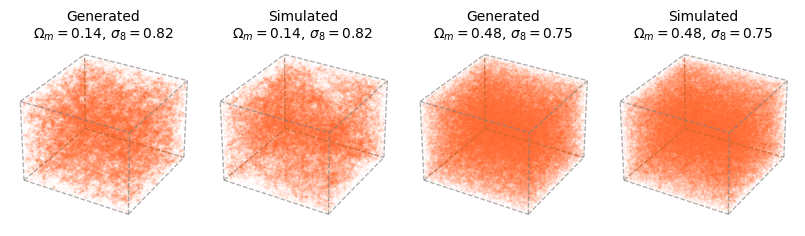}
    \caption{Comparison of two simulated and two generated cosmologies. The generated samples exhibit clustering patterns similar to those in the simulations.}
    \label{fig:viz_large_graph_194}
\end{figure}

\begin{figure}[h!]
    \centering
    \includegraphics{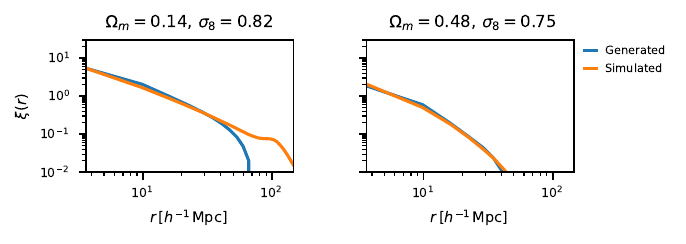}
    \caption{2PCF comparison between two simulated and generated cosmologies. Generated data faithfully reproduces 2PCF on smaller scales but deviates on larger scales.}
    \label{fig:2pcf-large-graph_194}
\end{figure}

Figure \ref{fig:viz_large_graph_194} shows that generated samples closely resemble the simulated matter distribution and follow the conditional cosmological dependence. The corresponding 2PCF in Figure \ref{fig:2pcf-large-graph_194} confirms that the large-scale structure is well reproduced, though clustering strength at large scales remains underestimated, consistent with observations on smaller graphs in Section~\ref{quali}.

\section{Limitations and future work}
This work demonstrates the potential of score-based generative models for generating large-scale cosmological structure, but several limitations remain. Score-based models introduce more hyperparameters for the inference process. Developing a theoretical framework for optimizing these settings in graph-based domains could speed up the generation process. Alternatively, flow matching presents a promising direction for future work, simplifying the inference process by avoiding Langevin dynamics. 

While our model captures variations in clustering strength across cosmologies, it struggles to reproduce long-range correlations. This difficulty in capturing long-range correlations, evident in the underestimation of the 2PCF at large scales in our generation experiments (Figure~\ref{fig:quali_tpcf} and Figure \ref{fig:2pcf-large-graph_194}), may stem from limitations inherent to vanilla graph neural networks, which are known to have difficulty modeling long-range dependence. Future investigations could explore multi-scale or hybrid GNN-Transformer architectures, aiming to leverage the GNN's efficiency for local interactions and the Transformer's capacity for global context, potentially mitigating the observed underestimation of large-scale clustering.

Finally, techniques such as exponential moving average could stabilize training and improve sample quality~\cite{song2020improved,jo2022score}, particularly for large graphs.

\section{Conclusion} 
We introduced a score-based generative model for generating large-scale structure conditioned on cosmological parameters. Our approach addresses key limitations of diffusion models by enforcing periodic boundary conditions during both training and inference, initializing sampling from a physically motivated uniform prior, and incorporating Euclidean symmetries. To effectively apply score-based models to large graphs, a topology-aware noise schedule was designed, guided by a quantitative measure of graph perturbation. We demonstrate the scalability of score-based generative modeling to graphs consisting of up to 600,000 nodes, marking the first such application in this context. Empirical results show that model performance is highly sensitive to the noise schedule and that our topology-aware scheduling significantly enhances generative quality. The model outperforms the diffusion-based baselines while offering substantial gains in computational efficiency and improved adherence to physical constraints.

Despite limitations in capturing long-range correlations, our model generates full cosmological samples with promising fidelity. The combination of physically grounded priors, symmetry preservation, and a novel topology-aware noise schedule constitutes a principled advancement in the design of generative models for cosmology, bringing such models closer to capturing the underlying physical processes that govern the evolution of the universe. Most notably, the introduction of the topology-aware noise schedule and the demonstrated scalability to realistic cosmological graph sizes constitute important developments toward machine learning-based simulators of matter clustering. Overall, these contributions move the field closer to developing viable, efficient, data-driven alternatives to computationally expensive $N$-body simulations.

\begin{credits}
\subsubsection{\ackname}
The project has been supported by the Dutch
national e-infrastructure with the support of SURF (grant no. EINF-10541). We thank SURF (www.surf.nl) for the support in using the National Supercomputer Snellius. 

\subsubsection{\discintname}
The authors have no competing interests to declare that are
relevant to the content of this article. 
\end{credits}

\bibliographystyle{splncs04}
\bibliography{mybibliography}

\appendix
\section{Implementation Details}
\label{appendix}
All models were trained for 1,000 epochs on 4 Nvidia H100 GPUs on the Dutch national supercomputer Snellius, using data parallelism with micro batch size 4. We used the AdamW optimizer with weight decay $1 \times 10^{-4}$, gradient clipping at 0.5, and a learning rate schedule with linear warm-up (epochs 0-100) from $1 \times 10^{-6}$ to $5 \times 10^{-4}$, followed by cosine decay. Most models use a noise schedule with 100 steps, $\sigma_L=0.001$ and $\sigma_1 = 0.2$; for ScoreEGNN (\( \Omega_m \), \( \sigma_8 \))** we used $\sigma_L=0.01$, $\sigma_1=1$. Langevin dynamics used $T=2$ and $\epsilon=3 \times 10^{-7}$ across all models. 

The equivariant GNN comprises 4 message-passing layers. Node and conditional features are projected to 16-D space and updated via 4-layer MLPs (128 hidden units, SiLU). Node messages are summed, added via residual connection, and normalized. Coordinates are updated via a 2-layer MLP (128 hidden units, SiLU, output 1). Attention scores, computed as in \cite{cuesta2023point}, scale messages using a 2-layer MLP (128 hidden units, output 1, SiLU). The non-equivariant GNN omits the coordinate model. Halo coordinates are node features, processed with initial embeddings via a 4-layer MLP (128 hidden units, output 16, SiLU), and final positions are decoded through another 4-layer MLP to 3D space.

\end{document}